\documentstyle[12pt]{article}
\begin{document}

 ~~\vspace{1cm}

\begin{center}

{\LARGE \sf
Relation between Dimension and Angular Momentum \\

\vspace{.2cm}

for Radially Symmetric Potential \\

\vspace{.2cm}

in $N$-dimensional Space}\\

\vspace{5mm} {\large  Zhao Wei-Qin$^{1,~2}$}

\vspace{11mm}

{\small \it 1. China Center of Advanced Science and Technology
(CCAST)}

{\small \it (World Lab.), P.O. Box 8730, Beijing 100080, China}

{\small \it 2. Institute of High Energy Physics, Chinese Academy
of Sciences,

P. O. Box 918(4-1), Beijing 100039, China}

\end{center}

\vspace{2cm}

\begin{abstract}

It is proved that when solving Schroedinger equations for radially
symmetric potentials the effect of higher dimensions on the radial
wave function is equivalent to the effect of higher angular
momenta in lower dimensional cases. This result is applied to
giving solutions for several radially symmetric potentials in
N-dimension.
\end{abstract}

\vspace{.5cm}
{\sf

 PACS{:~~03.65.-w,~~03.65.Ge}


}
\newpage
{\large \sf

There are more and more physical problems related to dimensions
higher than 3, which have attracted much attention recently[1]. In
the study of cosmology, group theory, many body problem,
supersymmetry, etc. multi-dimensional solutions are often
required. To solve the basic equation in quantum mechanics,
Schroedinger equation, for multi-dimensional problems serves very
well as the starting point for general discussions in any
multi-dimensional quantum problems.

Among the solutions of Schroedinger equation, the problems with
radially symmetric potential is of more interests. In fact, only
the solutions for a few radially symmetric potentials in 1-, 2-
and 3-dimensions are known. Much efforts have been paid to obtain
the solutions or to discuss their properties for higher
dimensional problems[1,2,3]. The purpose of this paper is to show
a very simple way to relate the solutions in 2- and 3- dimensional
problems to any higher dimensional cases for radially symmetric
potentials. The basic idea is that the dimension and the angular
momentum are related in a very simple way for radially symmetric
potentials. Based on this relation, the solutions with lower
angular momentum for higher dimensional problems can be expressed
by solutions with higher angular momentum for lower dimensional
problems, specially by 2- and 3-dimensional solutions.

In section 1 the general formula of the relation between
dimensions and angular momenta are given. It is pointed out that
the solutions for all odd-dimensional problems are related to
3-dimensional ones, while those for even-dimensional problems are
related to 2-dimensional solutions. In section 2, some examples
are presented to show how this formula can be applied to solving
Schroedinger equation for radially symmetrical potentials at
higher dimensions when the solutions for 2- and 3-dimension are
known. In Appendix a brief summary of the angular momentum in
$N$-dimensional space is given.

\newpage

\section*{\large \bf 1. Relation between Dimension and Angular Momentum}
\setcounter{section}{1} \setcounter{equation}{0}

Consider an N-dimensional Hamiltonian
\begin{eqnarray}\label{e2.1}
H = T + V(q)
\end{eqnarray}
where
\begin{eqnarray}\label{e2.2}
q =(q_1,~q_2,~\cdots,~q_N)
\end{eqnarray}
and
\begin{eqnarray}\label{e2.3}
T = -\frac{1}{2} \sum\limits_{i=1}^{N} \partial^2/\partial q_i^2 =
-\frac{1}{2} \nabla^2.
\end{eqnarray}
The corresponding Schroedinger equation is
\begin{eqnarray}\label{e2.4}
H\Psi(q)=E\Psi(q).
\end{eqnarray}

For a radially symmetric potential
\begin{eqnarray}\label{e2.5}
V(q)=V(r),
\end{eqnarray}
the radial part of the wave function can be solved as a one
dimensional problem by expressing Cartesian coordinates
$q_1,~q_2,~\cdots,~q_N$ in terms of the radial variable $r$ and
$(N-1)$ angular variables[2,4]
\begin{eqnarray}\label{e2.6}
\theta_1,~\theta_2,\cdots,~\theta_{N-2}~~{\sf and}~~\theta_{N-1}
\end{eqnarray}
through
\begin{eqnarray}\label{e2.7}
q_1&=&r \cos \theta_1,~~~~q_2=r \sin \theta_1 \cos
\theta_2,\nonumber\\
q_3&=&r \sin \theta_1 \sin \theta_2 \cos
\theta_3,~ \cdots, \\
q_{N-1}&=& r \sin \theta_1 \sin \theta_2 \cdots \sin \theta_{N-2}
\cos \theta_{N-1}\nonumber
\end{eqnarray}
and
\begin{eqnarray*}
~~~q_N~=~r \sin\theta_1 \sin \theta_2 \cdots \sin \theta_{N-2}
\sin \theta_{N-1}
\end{eqnarray*}
with
\begin{eqnarray}\label{e2.8}
0 &\leq& \theta_i< \pi ~~{\sf
for}~~i=1,~2,\cdots,~N-2\nonumber\\
{\sf and}~~~~~~~~~~~~~~&&\\
0&\leq& \theta_{N-1}\leq 2 \pi.\nonumber
\end{eqnarray}
The Laplacian operator can be expressed as[4]
\begin{eqnarray}\label{e2.9}
\nabla^2=\frac{1}{r^{2k}}~\frac{\partial}{\partial
r}(r^{2k}\frac{\partial}{\partial r}) -\frac{1}{r^2} {\cal
L}^2(N-1)
\end{eqnarray}
where
\begin{eqnarray*}
k=\frac{1}{2}(N-1)
\end{eqnarray*}
and  ${\cal L}^2(N-1)$ is the angular momentum operator. Its
definition is given in the Appendix. According to (A.5) in the
Appendix, the eigenvalues of each ${\cal L}^2(n)$ are
\begin{eqnarray}\label{e2.10}
l(l+n-1)
\end{eqnarray}
with $l=0,~1,~2,\cdots$. Thus, for a radially symmetric potential
the wave function can be written as
\begin{eqnarray}\label{e2.11}
\Psi(r,~\theta_1,~\theta_2,\cdots,~\theta_{N-1})={\cal R}(r)
\Theta (\theta_1,~\theta_2,\cdots,~\theta_{N-1})
\end{eqnarray}
Taking $l=l_1$ for the first equation of (A.7); i.e.,
\begin{eqnarray}\label{e2.12}
{\cal L}^2(N-1) \Theta = l(l+N-2) \Theta,
\end{eqnarray}
correspondingly, the radial part of the Schroedinger equation for
angular momentum $l$ is
\begin{eqnarray}\label{e2.13}
[-\frac{1}{2}\nabla_r^2(k) + \frac{1}{2r^2}l(l+N-2)+V(r)-E~]~{\cal
R}^l_k(r)=0
\end{eqnarray}
with
\begin{eqnarray}\label{e2.14}
 \nabla_r^2(k) = \frac{1}{r^{2k}}\frac{d}{dr}(r^{2k}\frac{d}{dr}).
\end{eqnarray}
Introducing
$$
{\cal R}^l_k(r)=\frac{1}{r^k}\psi(r) \eqno(1.15)
$$
$\psi(r)$ satisfies the following equation
$$
[-\frac{1}{2}\frac{d^2}{dr^2}+\frac{1}{2r^2}l(l+2k-1)+
\frac{1}{2r^2}k(k-1)+V(r)]\psi(r)=E\psi(r). \eqno(1.16)
$$
Considering
$$
l(l+2k-1)+k(k-1)=(l+k)(l+k-1) \eqno(1.17)
$$
we introduce
$$
K=l+k \eqno(1.18)
$$
and find that $\psi(r)$ is related only to the sum $K=l+k$, i.e.,
$\psi(r)=\psi_K(r)$, therefore, (1.16) can be written as
$$
[-\frac{1}{2}\frac{d^2}{dr^2}+\frac{1}{2r^2}K(K-1)+V(r)]\psi_K(r)=E\psi_K(r).
\eqno(1.19)
$$
Defining
$$
\nabla_r^2(K) = \frac{1}{r^{2K}}\frac{d}{dr}(r^{2K}\frac{d}{dr})
\eqno(1.20)
$$
and
$$
\overline{N}=2K+1=N+2l,\eqno(1.21)
$$
we can introduce the radial wave function
$$
{\cal R}_K(r)=\frac{1}{r^K}\psi_K(r) \eqno(1.22)
$$
satisfying
$$
[-\frac{1}{2}\nabla_r^2(K) +V(r)]~{\cal R}_K(r)=E~{\cal R}_K(r).
\eqno(1.23)
$$
This is just the $S$-state radial wave function for the same
radially symmetric potential in $\overline{N}$-dimension. It is
interesting to notice that for radially symmetric potential the
wave functions $\psi_K(r)$ for any angular momentum $l$ in
$N$-dimensional problem are the same as long as $l+k=K$ and
$N=2k+1$. For any combination $K=k+l$ the radial wave function
${\cal R}^l_k(r)$ with angular momentum $l$ in
$N(=2k+1)$-dimension can easily be obtained from this general
function $\psi_K(r)$ by
$$
{\cal R}^l_k(r)=\frac{1}{r^k}\psi_K(r)=r^l{\cal R}_K(r).
\eqno(1.24)
$$
(1.24) gives a very simple relation between radial wave functions
with different angular momenta $l$ and in different dimensions
$N(=2k+1)$ when $l+k=K$. Based on (1.24), for the same radially
symmetric potential they are related to the same $\psi_K(r)$
solved from (1.19) or to the same ${\cal R}_K(r)$ solved from
(1.23).

This relation at least has two kinds of applications: If the full
solutions for a potential in 2- and 3-dimension are known, such as
Coulomb potential or spherical harmonic oscillator, it is easy to
obtain the solutions for higher dimensions $N$ through the
relation (1.24). We will discuss this point in more details in the
following. On the other hand, if it is relatively easy to obtain
the solutions for $l=0$ in different dimensions $N$, either
analytically or approximately, the solutions for $l>0$ in lower
dimensions could be obtained by using (1.24). An application along
this direction is presented for $N$-dimensional Sombrero-shaped
potential in Ref.[4].

Now we discuss the first kind of application. Here it is necessary
to discuss two cases: When $N$ is odd, $k$ is an integer. In this
case, taking $K=L+1$ one reaches an equation in the 3-dimensional
space with angular momentum $L$
$$
[-\frac{1}{2}\frac{d^2}{dr^2}+\frac{1}{2r^2}L(L+1)+V(r)]\psi_K(r)=E\psi_K(r).
\eqno(1.25)
$$
Defining
$$
{\cal R}^L(r)=\frac{1}{r} \psi_K(r)=r^L{\cal R}_K(r)\eqno(1.26)
$$
(1.25) is equivalent to the following equation in the
3-dimensional space:
$$
[-\frac{1}{2}\nabla_r^2 +\frac{1}{2r^2}L(L+1) +V(r)]~{\cal
R}^L(r)=E~{\cal R}^L(r) \eqno(1.27)
$$
with $\nabla_r^2$ defined by
$$
 \nabla_r^2 = \frac{1}{r^{2}}\frac{d}{dr}(r^{2}\frac{d}{dr}).
\eqno(1.28)
$$
However, when $N$ is even, $k$ is a half integer. In this case,
taking $K=L+\frac{1}{2}$ one reaches an equation in 2-dimensional
space:
$$
[-\frac{1}{2}\frac{d^2}{dr^2}+\frac{1}{2r^2}(L^2-\frac{1}{4})+V(r)]\psi_K(r)=E\psi_K(r),
\eqno(1.29)
$$
where $L$ is the angular momentum in 2-dimensional space. Defining
$$
{\cal R}^L(r)=\frac{1}{\sqrt{r}} \psi_K(r)=r^L{\cal
R}_K(r)\eqno(1.30)
$$
(1.29) is equivalent to the following equation in 2-dimensional
space:
$$
[-\frac{1}{2}\nabla_r^2 +\frac{1}{2r^2}{L^2} +V(r)]~{\cal
R}^L(r)=E~{\cal R}^L(r) \eqno(1.31)
$$
with the corresponding $\nabla_r^2$ defined by
$$
 \nabla_r^2 = \frac{1}{r}\frac{d}{dr}(r\frac{d}{dr}).
\eqno(1.32)
$$
Therefore, if the solutions of (1.27) for any spherically
symmetric potential $V(r)$ in the 3-dimension case are known, the
solutions ${\cal R}_k^l(r)$ for the same potential in any odd
dimension $N=2k+1$ and at any angular momentum $l$ can easily be
derived via (1.26) and (1.24), taking $K=k+l=L+1$. On the other
hand, if the solutions of (1.31) for any spherically symmetric
potential $V(r)$ in the 2-dimension case are known, the solutions
${\cal R}_k^l(r)$ for the same potential in any even dimension
$N=2k+1$ and at any angular momentum $l$ can easily be derived via
(1.30) and (1.24), taking $K=k+l=L+\frac{1}{2}$.

In the following some examples are given to show how the relation
between dimensions and angular momenta is applied to solve high
dimensional problems based on the known solutions for the same
radial potentials in 2- or 3-dimensional cases.

\newpage

\section*{\large \bf 2. Some Examples}
\setcounter{section}{2} \setcounter{equation}{0}

{\bf i) $N$-dimensional Confined sphere}\\

Consider $N$-dimensional confined sphere. The potential is
expressed as
\begin{eqnarray}\label{e3.1}
V(r) =  \left\{\begin{array}{ccc}
0,~~~~~~~{\sf for}~~r <a\\
\infty,~~~~~~{\sf for}~~r>a.
\end{array}
\right.
\end{eqnarray}
The radial wave function for angular momentum $l$ is expressed by
${\cal R}^l_k(r)$, satisfying (1.13). Now we introduce $K=k+l$ and
defining
$$
\psi_K(r)=r^k{\cal R}^l_k(r)\eqno(2.2)
$$
according to (1.15). From (1.19) we know that $\psi_K(r)$
satisfies the following equation inside the confined sphere:
$$
[-\frac{1}{2}\frac{d^2}{dr^2}+\frac{1}{2r^2}K(K-1)]\psi_K(r)=E\psi_K(r)
\eqno(2.3)
$$
and outside the sphere the wave function is zero.

For odd dimensions $N=2k+1$, based on (1.26), defining the radial
wave function ${\cal R}^L(r)=\frac{1}{r}\psi_K(r)$ and introducing
angular momentum $L=K-1$ in 3-dimensional space, $\psi_K(r)$ also
satisfies the corresponding equation inside the confined sphere in
3-dimension:
$$
[-\frac{1}{2}\frac{d^2}{dr^2}+\frac{1}{2r^2}L(L+1)]\psi_K(r)=E\psi_K(r)
\eqno(2.4)
$$
and outside the sphere it is zero. Introducing
$$
q^2=2E\eqno(2.5)
$$
we have
$$
\psi_K''+[q^2-\frac{1}{r^2}L(L+1)]\psi_K=0.\eqno(2.6)
$$
The solution of (2.6) is the well known Bessel function[5]
$$
\psi_K(r)=c\sqrt{qr}J_{L+\frac{1}{2}}(qr)\eqno(2.7)
$$
and the corresponding radial wave function for the 3-dimensional
case is
$$
{\cal R}^L(r)=\frac{1}{r}\psi_K(r)\eqno(2.8)
$$
with eigenvalues
$$
E_{n_r,L}=\frac{1}{2}q^2_{n_r,L}~,\eqno(2.9)
$$
where $n_r$ is the radial quantum number.

Back to the odd $N$-dimensional case, remembering $K=L+1=k+l$, we
have
$$
{\cal R}^l_k(r)=\frac{1}{r^k}\psi_K(r)=\frac{1}{r^{k-1}}{\cal
R}^L(r)=\frac{c}{r^k}\sqrt{qr}J_{k+l-\frac{1}{2}}(qr) \eqno(2.10)
$$
and the corresponding eigenvalues are
$$
E_{n_r,k,l}=\frac{1}{2}q^2_{n_r,k+l-1}~.\eqno(2.11)
$$

Now we turn to even dimensions $N=2k+1$ with $k$ a half integer.
Based on (1.30), defining the radial wave function ${\cal
R}^L(r)=\frac{1}{\sqrt{r}}\psi_K(r)$ and introducing angular
momentum $L=K-\frac{1}{2}$ in 2-dimensional space, $\psi_K(r)$
satisfies the corresponding equation inside the confined sphere in
2-dimension:
$$
[-\frac{1}{2}\frac{d^2}{dr^2}+\frac{1}{2r^2}(L^2-\frac{1}{4})]\psi_K(r)=E\psi_K(r)
\eqno(2.12)
$$
and outside the sphere the wave function is also zero. Applying
the same $ q^2=2E$ as (2.5) we have
$$
\psi_K''+[q^2-\frac{1}{r^2}(L^2-\frac{1}{4})]\psi_K=0.\eqno(2.13)
$$
The solution of (2.13) is
$$
\psi_K(r)=c\sqrt{qr}J_{L}(qr)\eqno(2.14)
$$
and
$$
{\cal R}^L(r)=\frac{1}{\sqrt{r}}\psi_K(r).\eqno(2.15)
$$
with eigenvalues
$$
E_{n_r,L}=\frac{1}{2}q^2_{n_r,L-\frac{1}{2}}~.\eqno(2.16)
$$

Back to the even $N$-dimensional case, remembering
$K=L+\frac{1}{2}=k+l$, we have
$$
{\cal
R}^l_k(r)=\frac{1}{r^k}\psi_K(r)=\frac{1}{r^{k-\frac{1}{2}}}{\cal
R}^L(r)=\frac{c}{r^k}\sqrt{qr}J_{k+l-\frac{1}{2}}(qr) \eqno(2.17)
$$
and the corresponding eigenvalues are
$$
E_{n_r,k,l}=\frac{1}{2}q^2_{n_r,k+l-1}~.\eqno(2.18)
$$
\\

\noindent
{\bf ii) $N$-dimensional Harmonic Oscillator }\\

The potential of the $N$-dimensional harmonic oscillator is
expressed as
$$
V(r)=\frac{1}{2}\omega^2r^2.\eqno(2.19)
$$
The radial wave function for angular momentum $l$ is expressed by
${\cal R}^l_k(r)$, satisfying (1.13). As before, we introduce
$K=k+l$ and defining $ \psi_K(r)=r^k{\cal R}^l_k(r)$ as (2.2).
From (1.19) we know that $\psi_K(r)$ satisfies the following
equation
$$
[-\frac{1}{2}\frac{d^2}{dr^2}+\frac{1}{2r^2}K(K-1)+
\frac{1}{2}\omega^2r^2]\psi_K(r)=E\psi_K(r). \eqno(2.20)
$$

For odd dimensions $N=2k+1$, based on (1.26), defining the radial
wave function ${\cal R}^L(r)=\frac{1}{r}\psi_K(r)$ and introducing
angular momentum $L=K-1$ in 3-dimensional space, $\psi_K(r)$ also
satisfies the corresponding equation for harmonic oscillator in
3-dimension:
$$
[-\frac{1}{2}\frac{d^2}{dr^2}+\frac{1}{2r^2}L(L+1)+
\frac{1}{2}\omega^2r^2]\psi_K(r)=E\psi_K(r). \eqno(2.21)
$$
Defining $ q^2=2E$ we have
$$
\psi_K''+[q^2-\omega^2r^2-\frac{1}{r^2}L(L+1)]\psi_K=0.\eqno(2.22)
$$
The solution of (2.22) is well known:
$$
\psi_K(r)=r^{L+1}e^{-\frac{1}{2}\omega r^2}~
_1F_1(-n_r,~L+\frac{3}{2},~\omega r^2)\eqno(2.23)
$$
and
$$
{\cal R}^L(r)=\frac{1}{r}\psi_K(r)\eqno(2.24)
$$
with eigenvalues
$$
E_{n_r,L}=\omega (2n_r+L+\frac{3}{2}).\eqno(2.25)
$$
$_1F_1(-n_r,L+\frac{3}{2},\omega r^2)$ is the confluent
hypergeometric function[5] and $n_r$ the radial quantum number.

Back to the odd $N$-dimensional case, remembering $K=L+1=k+l$, we
have
$$
{\cal R}^l_k(r)=\frac{1}{r^k}\psi_K(r)=\frac{1}{r^{k-1}}{\cal
R}^L(r)=r^le^{-\frac{1}{2}\omega
r^2}~_1F_1(-n_r,~k+l+\frac{1}{2},~\omega r^2) \eqno(2.26)
$$
and the corresponding eigenvalues are
$$
E_{n_r,k,l}=\omega (2n_r+k+l+\frac{1}{2})\eqno(2.27)
$$
with $n_r$ the radial quantum number.

Now we turn to even dimensions $N=2k+1$ with $k$ a half integer.
Based on (1.30), defining the radial wave function ${\cal
R}^L(r)=\frac{1}{\sqrt{r}}\psi_K(r)$ and introducing angular
momentum $L=K-\frac{1}{2}$ in 2-dimensional space, $\psi_K(r)$
satisfies the corresponding equation for harmonic oscillator in
2-dimension:
$$
[-\frac{1}{2}\frac{d^2}{dr^2}+\frac{1}{2r^2}(L^2-\frac{1}{4})+
\frac{1}{2}\omega^2r^2]\psi_K(r)=E\psi_K(r). \eqno(2.28)
$$
Defining $ q^2=2E$ we have
$$
\psi_K''+[q^2-\omega^2r^2-\frac{1}{r^2}(L^2-\frac{1}{4})]\psi_K=0.\eqno(2.29)
$$
The solution is
$$
\psi_K(r)=r^{L+\frac{1}{2}}e^{-\frac{1}{2}\omega
r^2}~_1F_1(-n_r,~L+1,~\omega r^2)\eqno(2.30)
$$
and
$$
{\cal R}^L(r)=\frac{1}{\sqrt{r}}\psi_K(r)\eqno(2.31)
$$
with eigenvalues
$$
E_{n_r,L}=\omega (2n_r+L+1).\eqno(2.32)
$$

Back to the even $N$-dimensional case, remembering
$K=L+\frac{1}{2}=k+l$, we have
$$
{\cal
R}^l_k(r)=\frac{1}{r^k}\psi_K(r)=\frac{1}{r^{k-\frac{1}{2}}}{\cal
R}^L(r)=r^le^{-\frac{1}{2}\omega
r^2}~_1F_1(-n_r,~k+l+\frac{1}{2},~\omega r^2) \eqno(2.33)
$$
and the expression for the corresponding eigenvalues is the same
as (2.27), with $k$ a half integer here.\\

\noindent
{\bf iii) $N$-dimensional Coulomb Potential }\\

The $N$-dimensional Coulomb potential is expressed as
$$
V(r)=-\frac{Ze^2}{r}.\eqno(2.34)
$$
The radial wave function for angular momentum $l$ is expressed by
${\cal R}^l_k(r)$, satisfying (1.13). As before, we introduce
$K=k+l$ and defining $ \psi_K(r)=r^k{\cal R}^l_k(r)$ according to
(2.2). From (1.19) we know that $\psi_K(r)$ satisfies the
following equation
$$
[-\frac{1}{2}\frac{d^2}{dr^2}+\frac{1}{2r^2}K(K-1)-
\frac{Ze^2}{r}]\psi_K(r)=E\psi_K(r). \eqno(2.35)
$$

For odd dimensions $N=2k+1$, based on (1.26), defining the radial
wave function ${\cal R}^L(r)=\frac{1}{r}\psi_K(r)$ and introducing
angular momentum $L=K-1$ in 3-dimensional space, $\psi_K(r)$ also
satisfies the corresponding equation for Coulomb potential in
3-dimension:
$$
[-\frac{1}{2}\frac{d^2}{dr^2}+\frac{1}{2r^2}L(L+1)-
\frac{Ze^2}{r}]\psi_K(r)=E\psi_K(r). \eqno(2.36)
$$
Defining $\rho=\alpha r$, $\alpha^2=8|E|$ and
$\lambda=\frac{2Ze^2}{\alpha}=\frac{Ze^2}{\sqrt{2|E|}}$, the
solution of (2.36) is[5]
$$
\psi_K=\rho^{L+1}e^{-\frac{1}{2}\rho}{\cal
L}^{2L+1}_{n+L}(\rho)\eqno(2.37)
$$
and
$$
{\cal R}_{n,L}=\frac{1}{r}\psi_K\eqno(2.38)
$$
with eigenvalues
$$
E_n=\frac{Z^2e^4}{2n^2}.\eqno(2.39)
$$
${\cal L}^{2L+1}_{n+L}(\rho)$ is the Laguerre polynomial. The
parameters are fixed as
$$
\lambda=n=n_r+L+1~~~~~{\sf
and}~~~~\alpha_n=\frac{2Ze^2}{n}\eqno(2.40)
$$
with $n_r$ the radial quantum number.

Back to the odd $N$-dimensional case, remembering $K=L+1=k+l$, we
have
$$
{\cal R}^l_k=\frac{1}{r^k}\psi=\frac{1}{r^{k-1}}{\cal
R}^L=\rho^le^{-\frac{1}{2}\rho}{\cal
L}^{2(k+l)-1}_{n_r+2(k+l)-1}(\rho) \eqno(2.41)
$$
with the eigenvalues expressed by (2.39) and $n=n_r+k+l$.

Now we turn to even dimensions $N=2k+1$ with $k$ a half integer.
Based on (1.30), defining the radial wave function ${\cal
R}^L(r)=\frac{1}{\sqrt{r}}\psi_K(r)$ and introducing angular
momentum $L=K-\frac{1}{2}$ in 2-dimensional space, $\psi_K(r)$
satisfies the corresponding equation for Coulomb potential in
2-dimension:
$$
[-\frac{1}{2}\frac{d^2}{dr^2}+\frac{1}{2r^2}(L^2-\frac{1}{4})+
\frac{Ze^2}{r}]\psi_K(r)=E\psi_K(r). \eqno(2.42)
$$
As in the odd dimension case, defining $\rho=\alpha r$,
$\alpha^2=8|E|$ and $\lambda=\frac{2Ze^2}{\alpha}=
\frac{Ze^2}{\sqrt{2|E|}}$, the solution is
$$
\psi_K=\rho^{L+\frac{1}{2}}e^{-\frac{1}{2}\rho}{\cal
L}^{2L}_{n+L-\frac{1}{2}}(\rho)\eqno(2.43)
$$
and
$$
{\cal R}_{n,L}=\frac{1}{\sqrt{r}}\psi_K.\eqno(2.44)
$$
The eigenvalues are still expressed by (2.39), with
$n=n_r+L+\frac{1}{2}$.

Back to the even $N$-dimensional case, with $K=L+\frac{1}{2}=k+l$,
we have
$$
{\cal R}^l_k=\frac{1}{r^k}\psi_K=\frac{1}{r^{k-\frac{1}{2}}}{\cal
R}_{n,L}=\rho^le^{-\frac{1}{2}\rho}{\cal
L}^{2(k+l)-1}_{n_r+2(k+l)-1}(\rho) \eqno(2.45)
$$
with the eigenvalues still expressed by (2.39) and $n=n_r+k+l$.
All the obtained results for higher dimensions are consistent to
earlier works.[1,6]

From above 3 examples it is shown clearly that all the eigenstates
for radially symmetric $N$-dimensional potentials can be easily
obtained if the solutions for the 2- and 3-dimensional problems
with the same potential are known. Simply by substituting the
quantum numbers in the expressions of the wave functions and
eigenvalues according to the relation between the dimension and
the angular momentum, all the solutions for odd dimensions is
related to those of 3-dimensional case, while for the even
dimensions they are related to those of the 2-dimensional case. It
does not matter if the known solutions for 2- and 3-dimensions are
analytical or numerical, the only necessary condition is a clear
angular momentum dependence in the expressions. This method can
also be applied to approximate solutions under the same condition.

\newpage

\section*{{\bf Appendix A. Angular Momentum Operator}{\large [4]}}
\setcounter{section}{7} \setcounter{equation}{0}

Express the Cartesian coordinates $q_1,~q_2,~\cdots,~q_N$ in terms
of the radial variable $r$ and $(N-1)$ angular variables
$$
\theta_1,~\theta_2,\cdots,~\theta_{N-2}~~{\sf and}~~\theta_{N-1}
\eqno(A.1)$$ according to (1.7)-(1.8). Correspondingly, the line
elements are
\begin{eqnarray*}
~~~~~~~~~~~~~~~~~dr,~rd\theta_1,~r\sin \theta_1 d\theta_2,~r \sin
\theta_1
 \sin \theta_2 d \theta_3,~~~~~~~~~~~~~~~~~~~~~~~~(A.2)\\
 r \sin\theta_1 \sin \theta_2 \sin \theta_3 d\theta_4,\cdots,
 ~r \sin \theta_1 \sin\theta_2 \cdots \sin
 \theta_{N-2}d\theta_{N-1}.~~~~~~~~~~~~~~
\end{eqnarray*}
In the Laplacian operator given in (1.9)
$$
\nabla^2=\frac{1}{r^{2k}}~\frac{\partial}{\partial
r}(r^{2k}\frac{\partial}{\partial r}) -\frac{1}{r^2} {\cal
L}^2(N-1)
$$
the angular momentum operators ${\cal L}^2(N-1)$ are defined as
\begin{eqnarray*}
{\cal L}^2(N-1)&=&-\frac{1}{\sin^{N-2}
\theta_1}\frac{\partial}{\partial \theta_1}(\sin^{N-2}
\theta_1\frac{\partial}{\partial \theta_1})
+\frac{1}{\sin^2 \theta_1} {\cal L}^2(N-2) \nonumber\\
{\cal L}^2(N-2)&=&-\frac{1}{\sin^{N-3}
\theta_2}\frac{\partial}{\partial \theta_2}(\sin^{N-3}
\theta_2\frac{\partial}{\partial \theta_2})
+\frac{1}{\sin^2 \theta_2} {\cal L}^2(N-3) \nonumber\\
~~~~~\cdots&&\cdots~~~~~~~~~~~~~~~~~~~~~~~~~~~~~~~~~~~~~~~~~~~~~~~~~~~~~~~~~~~~~~~~~~(A.3)\\
{\cal L}^2(2)&=&-\frac{1}{\sin
\theta_{N-2}}\frac{\partial}{\partial \theta_{N-2}}(\sin
\theta_{N-2}\frac{\partial}{\partial \theta_{N-2}})
+\frac{1}{\sin^2 \theta_{N-2}} {\cal L}^2(1) \nonumber
\end{eqnarray*}
and
\begin{eqnarray*}
 {\cal L}^2(1)=-\frac{\partial^2}{\partial
 \theta^2_{N-1}}.~~~~~~~~~~~~~~~~~~~~~~~~~~~~~~~~~
\end{eqnarray*}
The square of the angular momentum operator on an $n$-sphere is
${\cal L}^2(n)$. The commutator between any two ${\cal L}^2(n)$
and ${\cal L}^2(m)$ is zero; i.e.,
$$
 [{\cal L}^2(n),~{\cal L}^2(m)]=0.
\eqno(A.4)$$

The eigenvalues of each ${\cal L}^2(n)$ are
$$
l(l+n-1) \eqno(A.5)
$$
with $l=0,~1,~2,\cdots$. The derivation of the eigenvalues of
${\cal L}^2(n)$ can be found in Appendix A of Ref.[4].

Thus, for a radially symmetric potential, the wave function can be
written as
$$
\psi(r,~\theta_1,~\theta_2,\cdots,~\theta_{N-1})={\cal R}(r)
\Theta (\theta_1,~\theta_2,\cdots,~\theta_{N-1}) \eqno(A.6)
$$
with
\begin{eqnarray*}
{\cal L}^2(N-1) \Theta &=&l_1(l_1+N-2) \Theta, \\
{\cal L}^2(N-2) \Theta&=&l_2(l_2+N-3) \Theta,
\end{eqnarray*}
$$
~~~~~~~~~~~\cdots \eqno(A.7)$$
\begin{eqnarray*}
{\cal L}^2(2) \Theta&=&l_{N-2}(l_{N-2}+1) \Theta
\end{eqnarray*}
and
\begin{eqnarray*}
{\cal L}^2(1) \Theta=l_{N-1}^2 \Theta.
\end{eqnarray*}

\section*{\bf Acknowledgement}

The author would like to thank Professor T. D. Lee for his
continuous guidance and instruction. This work is partly supported
by National Natural Science Foundation of China (NNSFC)
(No.10247001).

\section*{\bf References}
~~

[1] S. M. AL-Jaber and R. J. Lombard, J. Phys. A38(2005)4637 and

~~~~~~~the references therein.

[2] L. Chetouani and T. F. Hammann, J. Math. Phys. 27(1986)2944.

[3] S. M. Blinder, J. Math. Phys. 25(1984)905.

[4] R. Friedberg, T. D. Lee and W. Q. Zhao, preprint,
quant-ph/0510193.

[5] See, for example, S. Fluegge, "Practical Quantum Mechanics"

~~~~~~~Springer 1994.

[6] S. M. AL-Jaber, Int. Theoret. Phys. 37(1998)1289.

\end{document}